\documentclass[epj]{svjour}
\usepackage{amsbsy,amssymb,xspace,enumerate,color,mathptmx}
\makeatletter
\let\fref\@undefined
\let\Fref\@undefined
\makeatother
\expandafter\let\csname equation*\endcsname\relax
\expandafter\let\csname endequation*\endcsname\relax 
\usepackage{amsmath,graphicx}
\usepackage[plain]{fancyref}
\newcommand*{\bfrac}[2]{\genfrac{}{}{0pt}{}{#1}{#2}}
\newcommand{\upd}{\mathrm{d}}
\renewcommand{\vec}[1]{\mathbf{#1}}

\newcommand{\etal}{\textit{et.\,al.},\ }
\renewcommand{\eqref}[1]{(\ref{#1})}
\renewcommand{\bar}[1]{\overline{#1}}
\newcommand{\la}{\left \langle \left.}
\newcommand{\ra}{\right. \right \rangle}

\definecolor{mypurple}{RGB}{153,61,113}
\definecolor{myblue}{RGB}{63,61,153}
\definecolor{myokker}{RGB}{153,140,61}
\definecolor{mygreen}{RGB}{61,153,86}
\definecolor{mymarine}{RGB}{61,90,153}
\definecolor{mycyan}{RGB}{0,255,255}

\begin{document}
\title{Time correlations and persistence probability of a Brownian
  particle in a shear flow.}  \author{D Chakraborty} \institute{
  Max-Planck-Institut f\"ur Intelligente Systeme, Heisenberstr. 3,
  70569 Stuttgart, Germany, \and Institut f\"ur Theoretische und
  Angewandte Physik, Universit\"at Stuttgart, Pfaffenwaldring 57,
  70569 Stuttgart, Germany. } \abstract{In this article, results have
  been presented for the two-time correlation functions for a free and
  a harmonically confined Brownian particle in a simple shear
  flow. For a free Brownian particle, the motion along the direction
  of shear exhibit two distinct dynamics, with the
  mean-square-displacement being diffusive at short times while at
  late times scales as $t^3$. In contrast the cross-correlation $\la
  x(t) y(t) \ra $ scales quadratically for all times. In the case of a
  harmonically trapped Brownian particle, the mean-square-displacement
  exhibits a plateau determined by the strength of the confinement and
  the shear. Further, the analysis is extended to a chain of Brownian
  particles interacting via a harmonic and a bending
  potential. Finally, the persistence probability is constructed from
  the two-time correlation functions.} 

\maketitle
\section{Introduction}

The phenomenon of persistence has been well studied over the past
decade, both theoretically and experimentally
\cite{Majumdar1999,Ray2004}. Persistence is the probability that a
stochastic variable $x(t)$ retains a particular property up to the
observation time interval $t$ -- the property being either the sign of
the variable or the crossing of the origin
\cite{Ray2004,Majumdar1996,Bhattacharya2007}. The later definition of
$p(t)$ is synonymous with the survival probability of the stochastic
process, and the problem can also be approached using the backward
Fokker-Planck equation. For a wide class of non-equilibrium systems,
the asymptotic decay of $p(t)$ exhibits a power-law decay with a
non-trivial exponent $\theta$. This algebraic decay, and the exponent,
has been investigated for a wide class of non-interacting as well as
interacting systems including the overdamped Brownian walker in an
infinite \cite{Sire2000,Majumdar1999} and finite medium
\cite{Chakraborty2007}, diffusion equation with random initial
conditions \cite{Majumdar1996,Newman1998}, advection of a passive
scalar \cite{Chakraborty2009}, fluctuating interfaces
\cite{Krug1997,Toroczkai1999,Constantin2004}, critical dynamics
\cite{Majumdar1996b}, granular media \cite{Swift1999,Burkhardt2000},
disordered environments \cite{Fischer1998,Doussal1999,Chakraborty2008}
and polymer dynamics \cite{Bhattacharya2007}. The probability $p(t)$
acts as a dynamic probe, indicating how the system retains the memory
of its initial configuration as it evolves in time, and is also an
indirect test for the two-time correlation functions for a
non-stationary process. However, the estimation of the persistence
probability is notoriously difficult and an exact analytical
prediction for $p(t)$ at all times can be made only when the
stochastic process is Gaussian and Markovian, such as, for an
overdamped Brownian particle.

When the stochastic dynamics is a Gaussian Markovian process, the
non-stationary process $x(t)$ can be mapped to a stationary
Ornstein-Uhlenbeck process $\bar{X}(T)$, for which, the stationary
correlator $C(T)\equiv\la \bar{X}(T)\bar{X}(0) \ra $ is exponentially
decaying at all times. Asymptotically, the survival probability of the
stationary Ornstein-Uhlenbeck process is proportional to $C(T)$ and
$p(t)$ can be obtained by the inverse time transformation applied to
$\bar{X}$ \cite{Majumdar1999,Slepian1962}. For a wide class of
systems, the stationary correlator $C(T)$ is often non-exponential and
this straight-forward method can no longer be applied. Using the
classification of Slepian \cite{Slepian1962}, a correlator is of class
$\alpha$, if $C(T) \sim 1-\mathcal{O}(T^\alpha)$ in the limit of $T
\to 0$. The behavior of $C(T)$ near zero characterizes the density of
zero crossings $\rho$ of the stochastic process $\bar{X}$
\cite{Majumdar1996,Slepian1962,Rice1945}. When $\alpha=2$, the number of zero
crossings is finite and the Independent Interval Approximation can be
used to predict the exponent $\theta$ \cite{Majumdar1996}. If however,
$\alpha<2$, the stochastic process has infinte number of zero
crossings and a perturbative expansion gives a fairly good estimate
for $\theta$ \cite{Krug1997}.

The case of a free Brownian particle, in the overdamped limit, is
particularly simple and illustrates how the persistence probability
can be determined for a Gaussian Markov process \cite{Slepian1962}. We
define the persistence probability as the probability that the
particle has not crossed the origin during the observation time
interval $t$. Even though the system under consideration is very
simple, its application is abundant in quantitative science
\cite{Frey2005a}. The two-time correlation function for the
non-stationary process $x(t)=\int_0^t \eta(t') \upd t'$, where $\eta$
is a Gaussian stochastic noise, can be transformed into a stationary
Ornstein-Uhlenbeck process using the successive transformations
$\bar{X}\equiv x/\sqrt{\la x^2(t)\ra}$ and $T=\ln t$. The
stationary correlator of $\bar{X}$ is then given by $e^{-T/2}$, and
following Ref.\cite{Slepian1962}, the persistence probability in the
transformed variable decays exponentially -- $P(T) \sim e^{-T/2}$.
Using the reverse transformation, the decay of $p(t)$ in real time
becomes algebraic with an exponent $\theta=1/2$. 

In this article, we investigate the persistence probability of a free
and confined Brownian particle in a shear flow and that of an
interacting chain of Brownian particles. The work is motivated on one
hand by the extensive theoretical
\cite{SanMiguel1979,Taylor1953,Yoshishige1996,Kienle2011,Bammert2010}
and experimental \cite{Orihara2011} study of Brownian motion in a
linear shear flow and their importance in microfluidic applications,
and on the other hand, due to the rich dynamics exhibited by such
systems as a result of an interplay of thermal fluctuations and the
imposed velocity gradient
\cite{Perkins1997,Groisman2001,Groisman2000}. The investigation of the
survival probability for a free Brownian particle in a deterministic
flow field $v(x)$ has been previously investigated
\cite{Bray2006,Bray2005,Bray2004}, and is known that for all odd
functions of $v(x)$ the survival probability decays as
$t^{-1/4}$. However, all of these approaches use the Fokker-Planck
equation to determine the survival probability. In the present work,
we emphasize on the two-time correlation function and present explicit
results for them for the case of a free and harmonically confined
Brownian particle in a transverse flow field. In the case of a chain
of Brownian particles harmonically bounded to its nearest neighbors,
we impose an additional bending potential to mimic the case of a
semi-flexible polymer chain.

The rest of the article is organized as follows: the model system is
introduced in \fref{sec:simple_shear}. The relevant results for the
two-time correlation functions and the persistence probability for a
free Brownian particle is presented in \fref{sec:simple_shear}, and
for a confined Brownian particle, in \fref{sec:harmonically_bound}.
We extend the analysis presented in the previous sections to a chain
of interacting Brownian particles and present the results in
\fref{sec:interacting_chain}.

\section{Brownian particle in a shear flow}
\label{sec:simple_shear}
The stochastic dynamics of a Brownian particle can be looked at from
different levels of coarse-graining. Typically, if the measurement
time intervals of the position and momenta of a colloid are well
separated, the Markovian Langevin equation in the overdamped limit
faithfully reproduces the dynamics. In reality, however, the dynamics
of a Brownian particle exhibit a far more rich physics. As opposed to
the Markovian Langevin equation, characterized by a single relaxation
time, the actual dynamics of Brownian particle is characterized by a
set of time-scales determined by sound propagation, vorticity
diffusion and the inertia of the colloid. As a consequence, the motion
of the Brownian particle no longer remains Markovian and the
generalized Markovian equation with a memory dependent friction and a
correlated noise is used to describe the dynamics. With this
increasing level of complexity, an analytically tractable result
becomes difficult and one has to resort to a numerical integration of
the equations of motion. In the following, however, we consider the
overdamped Langevin equation to describe the dynamics, neglecting the
effects of solvent and the inertia of the Brownian particle.

We consider the motion of a Brownian particle, with a unit mass, in an
unbounded solvent moving in a two-dimensional planar
geometry. Following \cite{SanMiguel1979}, we consider a stationary
distribution of velocity,
\begin{equation}
  \label{eq:velprf}
  \vec{u}=(0,ax).
\end{equation}
The force on the Brownian particle, due to the imposed flow is given
by $\vec{F}=-\zeta(\vec{v}-\vec{u})$, where $\vec{v}$ is the
instantaneous velocity of the particle and $\zeta$ is the Stokes's
friction on the colloid. The Langevin equation for the position of a
colloid $\vec r \equiv (x,y)$, in the overdamped limit, takes the form
\begin{equation}
  \label{eq:langevin}
  \frac{\upd x}{\upd t}=\eta_{x}(t) \quad \quad \quad  
  \textrm{and} \quad \quad \quad \frac{\upd y}{\upd t}=a x+
  \eta_{y}(t),
\end{equation}
with $\boldsymbol{\eta}\equiv(\eta_x, \eta_y)$ as a Gaussian white
noise with correlations
\begin{equation}
  \label{eq:noisecorr}
  \la \boldsymbol{\eta}(t)\ra=0 \quad \quad \quad \textrm{and}
  \quad \quad \quad
  \la \boldsymbol{\eta}(t)\otimes\boldsymbol{\eta}(t') \ra=
  2 D \mathbf{I}\delta(t-t'),
\end{equation}
where $\mathbf{I}$ is the identity matrix and $\otimes$ denote the
outer product of a vector quantity. The strength of the noise
correlations is given by the diffusion constant $D=k_{\rm B}T/\zeta$.
The two-time correlation functions from \fref{eq:langevin} becomes,
\begin{equation}
  \label{eq:twotimex}
  \la x(t_1) x(t_2)\ra=\int_0^{t_1} \upd t_1'\int_0^{t_2} \upd
  t_2'  \la \eta_x(t_1')\eta_x(t_2') \ra
\end{equation}
and
\begin{multline}
  \label{eq:twotimey}
  \la y(t_1) y(t_2)\ra=\int_0^{t_1} \upd t_1'\int_0^{t_2} \upd
  t_2'\;  \la \eta_y(t_1')\eta_y(t_2') \ra\\
  +a^2 \int_0^{t_1} \upd t_1'\int_0^{t_2} \upd t_2'  \;\la
  x(t_1')x(t_2') \ra.
\end{multline}
Assuming $t_1>t_2$, \fref{eq:twotimex} yields $\la x(t_1)
x(t_2)\ra=2Dt_2$ and
\begin{equation}
  \label{eq:twotimey1}
  \la y(t_1) y(t_2)\ra= 2 D t_2+a^2\int_0^{t_1} \upd t_1'\int_0^{t_2} \upd t_2'  \min(t_1',t_2').
\end{equation}
The integral on the right-hand-side of \fref{eq:twotimey1} evaluates
to 
\begin{equation}
  \label{eq:twotimey2}
  \la y(t_1) y(t_2)\ra= 2 D t_2+a^2 D \left(t_1 t_2^2 -
  \frac{t_2^3}{3}\right).
\end{equation}
Similarly, the cross-correlation functions $\la x(t_1)
y(t_2)\ra$ and $\la y(t_1) x(t_2) \ra$ can be constructed
as,
\begin{multline}
  \label{eq:cross_corr_xy}
   \la x(t_1) y(t_2)\ra=a\int_0^{t_1} \upd t_1' \int_0^{t_2}
  \upd t'_2 \;\, \la \eta_x(t'_1)x(t_2') \ra\\
  + \int_0^{t_1} \upd t_1' \int_0^{t_2} \upd t'_2 \;\, \la
  \eta_x(t'_1)\eta_y(t_2') \ra,
\end{multline}
and
\begin{multline}
  \label{eq:cross_corr_yx}
   \la y(t_1) x(t_2)\ra=a\int_0^{t_1} \upd t_1' \int_0^{t_2}
  \upd t'_2 \;\, \la x(t'_1) \eta_x(t'_2) \ra\\
  + \int_0^{t_1} \upd t_1' \int_0^{t_2} \upd t'_2 \;\, \la
  \eta_x(t'_1)\eta_y(t_2') \ra.
\end{multline}
The second terms in both of the above equation are zero due to the
choice of the noise correlations in \fref{eq:noisecorr}.
Further, using the solution  $x(t)=\int_0^t \upd t'\eta_x(t')$, the
correlation function in \fref{eq:cross_corr_xy} becomes,
\begin{align}
  \label{eq:corr_xy_1}
  \la x(t_1) y(t_2)\ra=a \int_0^{t_2} \upd t'_2 \int_0^{t_1}
  \upd t_1' \int_0^{t'_2} \upd t''_2\;\, \la
  \eta_x(t'_1)\eta(t''_2) \ra= D a t_2^2,
\end{align}
and the correlation function in \fref{eq:cross_corr_yx}
evaluates to,
\begin{equation}
  \label{eq:corr_yx_1}
 \begin{split}
  \la y(t_1) x(t_2)\ra&=a\int_0^{t_1} \upd t_1' \int_0^{t_2}
  \upd t'_2 \int_0^{t'_1} \upd t''_1 \;\, \la \eta_x(t''_1) 
  \eta_x(t'_2) \ra\\ &= 2Da \int_0^{t_1} \upd t'_1
  \min(t_2,t'_1)=2Da\left(t_1 t_2 -\frac{t_2^2}{2}\right).
\end{split}
\end{equation}

\begin{figure*}
\hfill
  \includegraphics[width=0.45\linewidth]{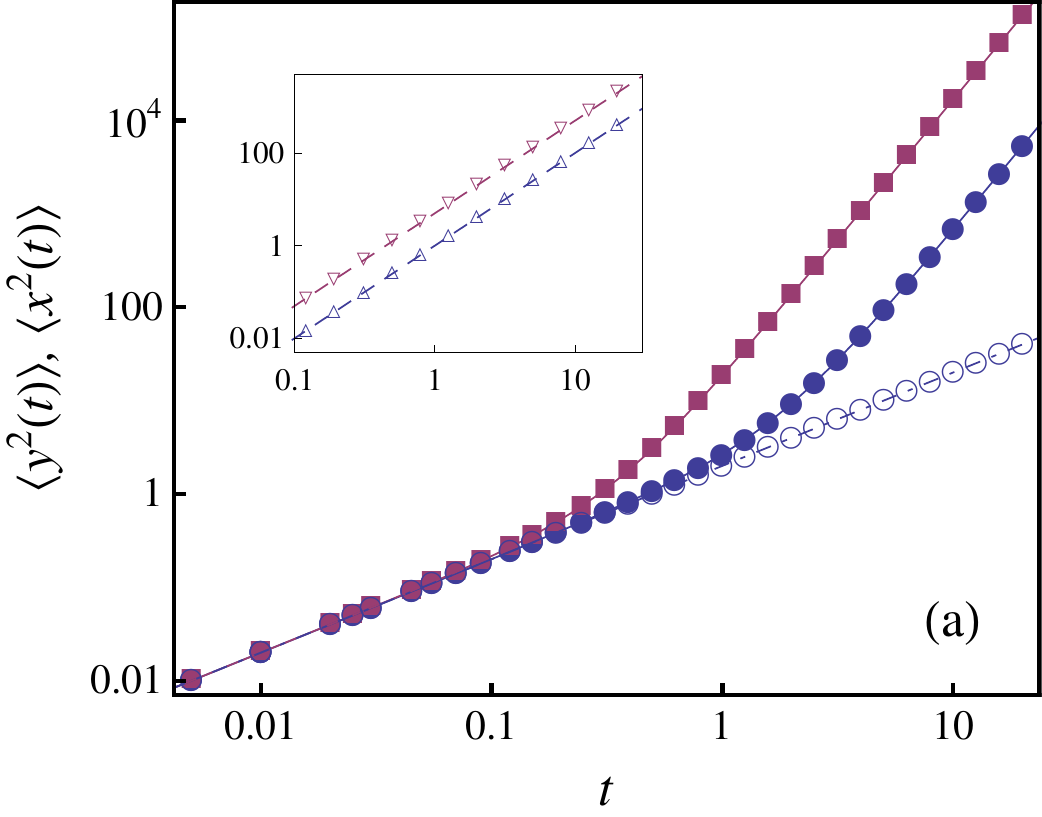}
\hfill
  \includegraphics[width=0.45\linewidth]{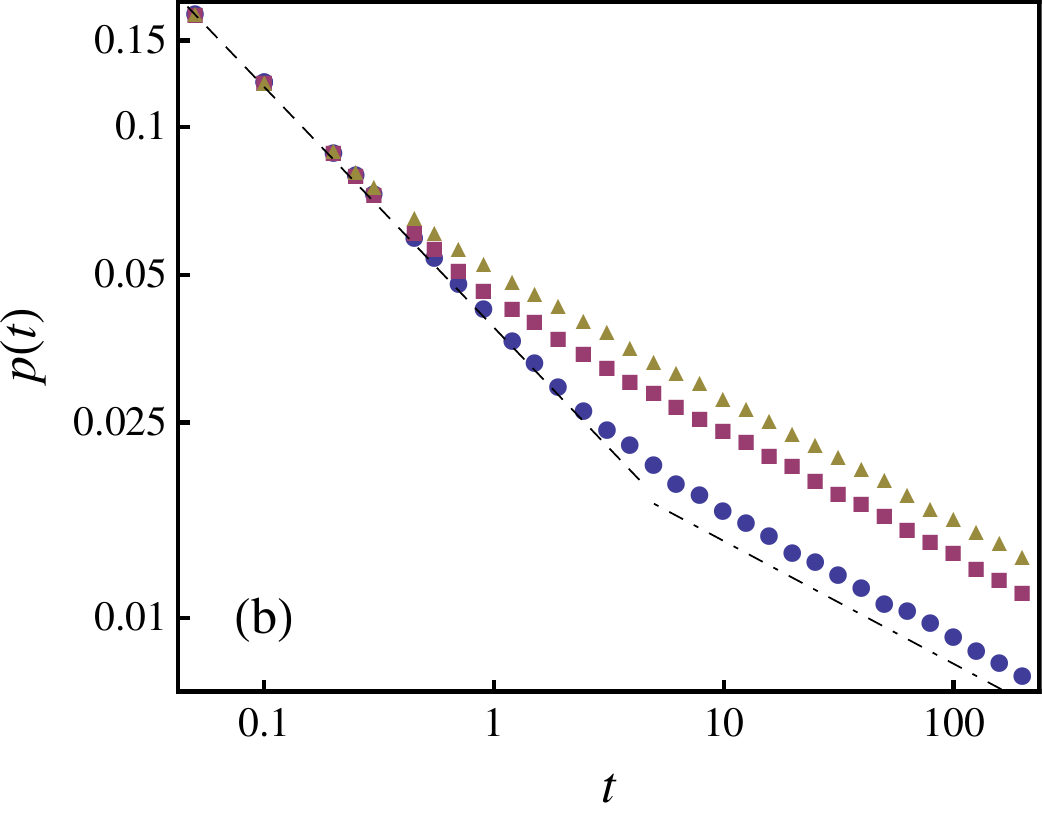}
  \caption{\textbf{(a)} Mean-square displacement of a Brownian
    particle in a simple shear along the $y$ direction for $a=1.00$
    ({\color{myblue} {\large $\bullet$}}) and $5.00$
    ({\color{mypurple} $\blacksquare$}). The corresponding mean-square
    displacement along the $x$-direction is shown for $a=1.00$
    ({\Large \color{myblue} $\circ$}) only. The solid lines are the
    plots of \fref{eq:msdy} for the respective values of $a$ and the
    dot-dashed line is the plot of $\la x^2(t)\ra= 2 D
    t$. \emph{\textbf{Inset:}} Plot of $\la x(t) y(t)\ra$ for $a=1.00$
    ({\color{myblue} { $\vartriangle$}}) and $5.00$ ({\color{mypurple}
      $\triangledown$}). The dashed line are the plots of
    \fref{eq:xyt}. \textbf{(b)} Persistence probability of a Brownian
    particle in a simple shear for ({\color{myblue} {\large
        $\bullet$}}), $5.00$ ({\color{mypurple} $\blacksquare$}) and
    $10.0$ ({\color{myokker} $\blacktriangle$}). The dashed line is
    plot of $t^{-1/2}$ corresponding to that of a free Brownian
    particle, while the dot dashed line is a plot of $t^{-1/4}$. }
  \label{fig:msdplot_freebm}
\end{figure*}

Using the two-time correlation functions evaluated above, we can
construct the corresponding mean-square displacements by simply
substituting $t_2=t_1=t$. Along the $x$-direction the motion remains
purely diffusive, while the mean-square-displacement of the particle
along the $y$-direction becomes,
\begin{equation}
  \label{eq:msdy}
  \la y^2(t)\ra= 2 D t+\frac{2}{3}a^2 D t^3.
\end{equation}
For short times, $t<<a^{-1}$, the motion of the Brownian particle
along the $y$-direction is purely diffusive, while in the asymptotic
regime of $t>>a^{-1}$ the mean-square-displacement scales as $t^3$ --
similar to that of a randomly accelerated particle.

The cross-correlation function $\la x(t) y(t) \ra$, however,
has two contributions and adding up \fref{eq:corr_xy_1} and
\fref{eq:corr_yx_1} gives,
\begin{equation}
  \label{eq:xyt}
  \la x(t) y(t) \ra= Da t^2.
\end{equation}

Our aim is to extract the persistence probability of the Brownian
particle using the correlation function defined in
\fref{eq:twotimey2}.  To this end, we define it as the probability
that the $y$-component of the particle position has not crossed
zero. The fundamental idea is to map the non-stationary process $y(t)$
to a stationary Ornstein-Uhlenbeck (O-U) process $\overline{Y}$. The
stationary correlator $C(T)$ for $\overline{Y}$ decays exponentially
for all times and the persistence probability can be shown to decay as
$P(T)\sim \frac{2}{\pi}\sin^{-1}[C(T)]$
\cite{Slepian1962,Majumdar2001}. The mapping to the stationary O-U
process is done using two transformations, first normalizing the
stochastic variable by the root-mean-square displacement and then a
suitable transformation in time. For the correlation function in
\fref{eq:twotimey2}, the transformation $\overline{Y}=y(t)/\sqrt{\la
  y^2(t)\ra}$ yields,
\begin{equation}
  \label{eq:Y1Y2}
  \la \overline{Y}(t_1) \overline{Y}(t_2) \ra= \dfrac{2Dt_2 +a^2D
    \left(t_1 t_2^2 -\dfrac{t_2^3}{3} \right)}{\left(2Dt_1+
     \frac{2}{3} a^2Dt_1^3\right)^{1/2}\left(2Dt_1+
     \frac{2}{3} a^2Dt_1^3\right)^{1/2}}
\end{equation}
A suitable time transformation which converts \fref{eq:Y1Y2} to a
stationary process for all times is non-trivial. However, we can
extract persistence probability by considering the limiting behaviors
of \fref{eq:Y1Y2}. As noted above, for $t\ll a^{-1}$, the motion of the
colloid is purely diffusive and neglecting the $\mathcal{O}(t^3)$
terms gives,
\begin{equation}
  \label{eq:Y1Y2_short}
  \la \overline{Y}(t_1) \overline{Y}(t_2) \ra=\sqrt{\dfrac{t_2}{t_1}}.
\end{equation}
 In the opposite limit of $t \gg a^{-1}$, the linear term can be
 neglected and the correlation function for the normalized variable
 becomes,
 \begin{equation}
   \label{eq:Y1Y2_long}
   \la \overline{Y}(t_1) \overline{Y}(t_2) \ra= \frac{3}{2}
   \left(\frac{t_2}{t_1}\right)^{1/2} - \frac{1}{2}
   \left(\frac{t_2}{t_1}\right)^{3/2}
 \end{equation}
 Using the time transformation $e^T=t$, \fref{eq:Y1Y2_short} and
 \fref{eq:Y1Y2_long} are transformed to a Gaussian stationary process
 with correlations,
 \begin{equation}
   \label{eq:corr_Y}
   C(T)=\begin{cases} e^{-T/2} & \mbox{for } t \ll a^{-1}\\
  \frac{3}{2} e^{-T/2} -\frac{1}{2} e^{-3T/2} & \mbox{for } t \gg a^{-1}.
  \end{cases}
 \end{equation}
 Since the stationary correlator for $t<a^{-1}$ is exponentially
 decaying, the persistence probability in the transformed variable $T$
 is given by $P(T) \sim e^{-T/2}$ and in real time $p(t) \sim
 t^{-1/2}$. In the asymptotic regime, the correlation function
 corresponds to that of a randomly accelerated particle and the
 persistence probability is known from the works of Sinai and
 Burkhardt \cite{Sinai1992,Burkhardt2000} to decay as $p(t) \sim t^{-1/4}$.

 To validate the results presented above, numerical simulations were
 done by integrating \fref{eq:langevin} using the Euler scheme with an
 integration time step of $\upd t =0.001$. From this numerical
 integration, the position of the particle in the subsequent time
 steps were measured and the mean-square-displacement along and
 perpendicular to the direction of the applied shear, as well as the
 cross-correlation functions were computed. The results of the
 numerical simulation is presented in the left panel of
 \fref{fig:msdplot_freebm} and compared to the analytical results of
 \fref{eq:msdy} and \fref{eq:xyt}. The measured data were averaged
 over $10^5$ independent configurations. In all of these simulation
 runs, the particle was started in the neighborhood of zero, so that
 the sign of $y(0)$ is defined and subsequently the sign of $y(t)$ was
 followed. The fraction of particles for which the sign of $y(t)$ did
 not change quantified the persistence probability. The measured
 persistence probability for the Brownian particle in the direction of
 the applied shear is shown in the right panel of
 \fref{fig:msdplot_freebm}. The measured probability exhibits two
 distinct regimes of decay, for $t<<a^{-1}$ a decay of $t^{-1/2}$ and
 for $t>> a^{-1}$ a decay of $t^{-1/4}$.

\section{Harmonically confined Brownian particle in a shear flow}
\label{sec:harmonically_bound}
In this section we study the dynamics of a tracer in a simple shear
flow which is also confined by a harmonic potential. The harmonic
confinement occurs naturally in the experiments when a tracking of a
tracer is done using an optical tweezers. The equations of motion for
the position of a colloid which is subjected to a harmonic confinement
$U(r)=\frac{1}{2}\kappa r^2$ and the stationary velocity profile of
\fref{eq:velprf} are,
\begin{equation}
  \label{eq:langevin_harmonic}
  \frac{\upd x}{\upd t} =-\kappa x +\eta_x(t) \;\; \; \textrm{and} \;\;\;
  \frac{\upd y}{\upd t} =-\kappa y + a x + \eta_y(t),
\end{equation}
together with the noise correlations of \fref{eq:noisecorr}. The time
evolution of the coordinates is given by,
\begin{equation}
  \label{eq:xt_yt}
\begin{split}
  x(t)=\int_0^t \upd t' e^{-\kappa (t-t')} \eta_x(t') \;\;
  \mbox{and}\;\; \\
  y(t)=\int_0^t \upd t' e^{-\kappa (t-t')}
  \left[x(t')+\eta_y(t')\right].
\end{split}
\end{equation}
Assuming that $t_1>t_2$, the two-time correlation function for
the variable $x(t)$ is straightforward and gives,
\begin{equation}
  \label{eq:two_time_hc_x}
  \la
  x(t_1)x(t_2)\ra=\frac{D}{\kappa}\left[e^{-\kappa(t_1-t_2)}-
  e^{-\kappa(t_1+t_2)}\right]
\end{equation}
while for $y(t)$ we have,
\begin{equation}
  \label{eq:two_time_hc_y}
\begin{split}
  \la y(t_1) y(t_2) \ra=\frac{a^2D}{\kappa}e^{-\kappa(t_1+t_2)}
  \int_0^{t_1} \upd t'_1\int_0^{t_2} \upd t'_2\;\,e^{\kappa(t'_1+t'_2)} 
 \left[ e^{-\kappa|t'_1-t'_2|} - \right.\\
  \left. e^{-\kappa(t'_1+t'_2)} \right]
+\int_0^{t_1} \upd t'_1\int_0^{t_2}
 \upd t'_2\;\,\la\eta_y(t'_1) \eta_y(t'_2) \ra.
\end{split}
\end{equation}
The integral over $t'_1$ and $t'_2$ yields,
\begin{multline}
  \label{eq:two_time_hc_y1}
  \la y(t_1) y(t_2) \ra=\frac{a^2D}{2\kappa^3}
  \left[\left(e^{-\kappa(t_1-t_2)}-e^{-\kappa(t_1+t_2)}\right) \right.\\
  -\kappa\left((t_1+t_2)e^{-\kappa(t_1+t_2)}
    -(t_1-t_2)e^{-\kappa(t_1-t_2)}\right) \\
  - \left. 2 \kappa^2\; t_1 t_2
    e^{-\kappa(t_1+t_2)}\right]+\frac{D}{\kappa}\left[
    e^{-\kappa(t_1-t_2)} - e^{-\kappa(t_1+t_2)}\right]
\end{multline}
The mean-square displacement for $y(t)$ immediately follows from
\fref{eq:two_time_hc_y1} via a substitution $t_1=t_2=t$,
\begin{multline}
  \label{eq:msd_hc_y}
  \la y^2(t)\ra=\frac{a^2D}{2\kappa^3}
  \left[\left(1-e^{-2\kappa t}\right)+2\kappa t e^{-2 \kappa
      t}-2\kappa^2 t^2 e^{-2 \kappa t} \right]\\
  +\frac{D}{\kappa}\left[1-e^{-2 \kappa t} \right].
\end{multline}
\begin{figure}
  \centering
  \includegraphics[width=\linewidth]{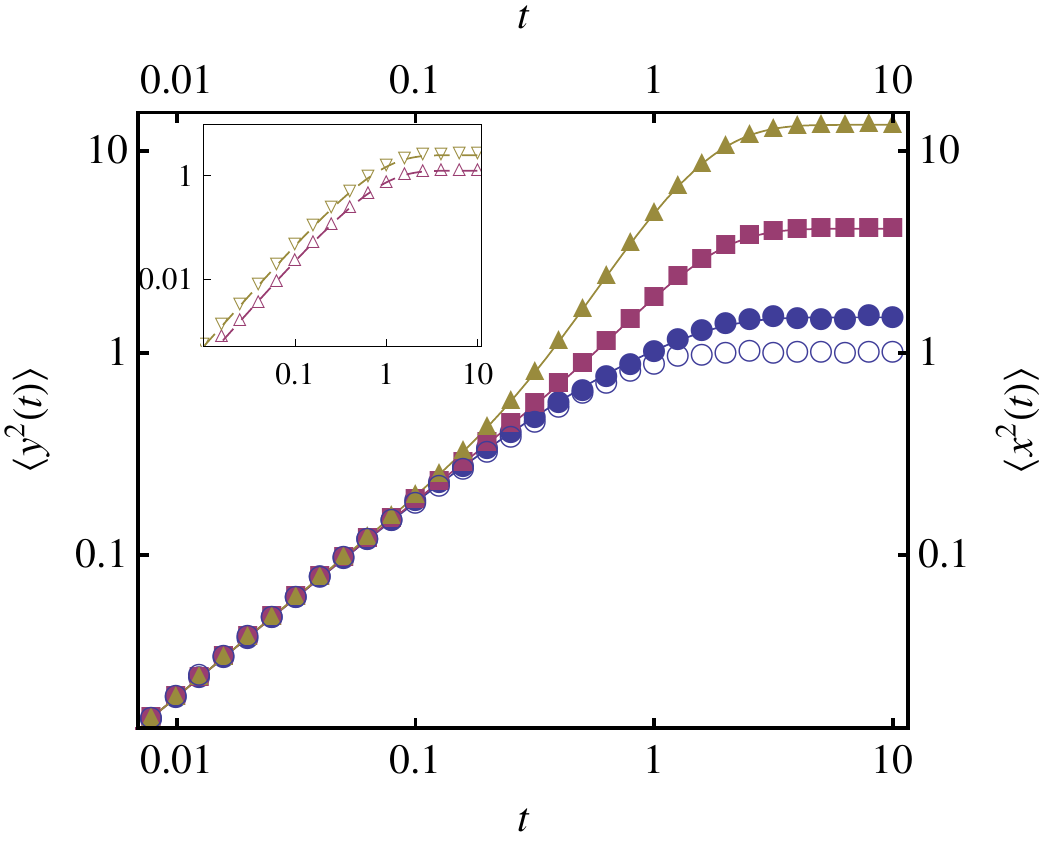}
  \caption{Mean-square displacement of a Brownian particle in a simple
    shear along the $y$ direction for $a=1.00$ ({\color{myblue}
      {\large $\bullet$}}), $2.50$ ({\color{mypurple}
      $\blacksquare$}) and $5.00$ ({\color{myokker}
      $\blacktriangle$}). The corresponding mean-square displacement
    along the $x$-direction is shown for $a=1.00$ ({\Large
      \color{myblue} $\circ$}) only. The solid lines are the plots of
    \fref{eq:msd_hc_y} for the respective values of $a$ and the dot-dashed
    line is the plot of $\la x^2(t)\ra= 2 D
    t$. \emph{\textbf{Inset:}} Plot of $\la x(t) y(t)\ra$ for
    $a=2.50$ ({\color{mypurple} { $\vartriangle$}}) and $5.00$
    ({\color{myokker} $\triangledown$}). The dashed line are the plots
    of \fref{eq:hc_xy_t}.}
  \label{fig:msd_hc}
\end{figure}
Further, the cross-correlation functions 
$\la x(t_1)y(t_2)\ra$ and \\
$ \la y(t_1)x(t_2) \ra$ is given by,
\begin{multline}
  \label{eq:corr_hc_xy}
  \la x(t_1) y(t_2)\ra= \int_0^{t_2} \upd t'_2\;\; e^{-\kappa
    (t_2-t'_2)} \la x(t_1) x(t'_2)
  \ra\\
  =\frac{a D}{2 \kappa^2} \left[e^{-\kappa (t_1 -t_2)} -
    e^{-\kappa (t_1+t_2)}-2\kappa t_2e^{-\kappa(t_1+t_2)}\right]
\end{multline}
\begin{multline}
  \label{eq:corr_hc_yx}
  \la y(t_1) x(t_2)\ra= \int_0^{t_1} \upd t'_1\;\;
  e^{-\kappa(t_1-t'_1)}\la x(t'_1) x(t_2)\ra\\
 =\frac{a D}{ \kappa}\int_0^{t_1} \upd t'_1\;\;e^{-\kappa(t_1-t'_1)} 
\left[e^{-\kappa |t'_1 -t_2|} -e^{-\kappa (t'_1+t_2)}\right]
\end{multline}
The integral over $t'_1$ in \fref{eq:corr_hc_yx} gives,
\begin{multline}
  \label{eq:corr_hc_yx1}
  \la y(t_1) x(t_2)\ra= \frac{a D}{2 \kappa^2}
   \left[ e^{-\kappa (t_1 -t_2)} - e^{-\kappa (t_1+t_2)} 
   + 2 \kappa (t_1 - t_2) e^{-\kappa(t_1-t_2)}\right.\\
    \left. -2\kappa t_1 e^{-\kappa (t_1+t_2)}\right]
\end{multline}
The cross-correlation function $\la x(t) y(t) \ra$ is obtained
by adding \fref{eq:corr_hc_xy} and \fref{eq:corr_hc_yx1} and the
substitution $t_1=t_2=t$ which gives,
\begin{equation}
  \label{eq:hc_xy_t}
  \la x(t) y(t) \ra=\frac{a D}{2 \kappa^2}\left[ 1-e^{-2
      \kappa t} - 2 \kappa t e^{-2 \kappa t} \right]
\end{equation}

A Taylor expansion of \fref{eq:msd_hc_y} for $t<\kappa^{-1}$, shows
that the dynamics at short times scales as $2Dt+(2/3) a^2 D t^3$,
while in the asymptotic regime, the mean-square displacement saturates
to a value $D/\kappa+a^2D/\kappa^3$. To determine the persistence
probability, we follow the steps outlined in the previous
section. However, an expression for the persistence probability valid
at all times is once again not feasible. Even in the asymptotic
regime, a suitable time transformation does not exist which converts
the process $y(t)$ to a Gaussian stationary process. Progress can only
be made in the time domain which is smaller than $\kappa^{-1}$, where
the two-time correlation function is identical to \fref{eq:twotimey2}.
Accordingly, the persistence probability shows an initial decay of
$t^{-1/2}$, followed by a decay of $t^{-1/4}$.
The mean-square-displacement of the harmonically confined Brownian
particle in a simple shear flow is shown in \fref{fig:msd_hc}. Once
again, the integration of \fref{eq:langevin_harmonic} was carried out
using the Euler scheme with an integration time step of $\upd
t=0.001$. The measured mean-square-displacement is compared to the
analytical prediction of \fref{eq:msd_hc_y} in the main figure, while
the inset depicts the cross-correlation function $\la x(t) y(t) \ra $
for two different values of $a$, defined in \fref{eq:velprf}.

\section{Interacting chain of Brownian particles}
\label{sec:interacting_chain}

\begin{figure}
  \centering
  \includegraphics[width=\linewidth]{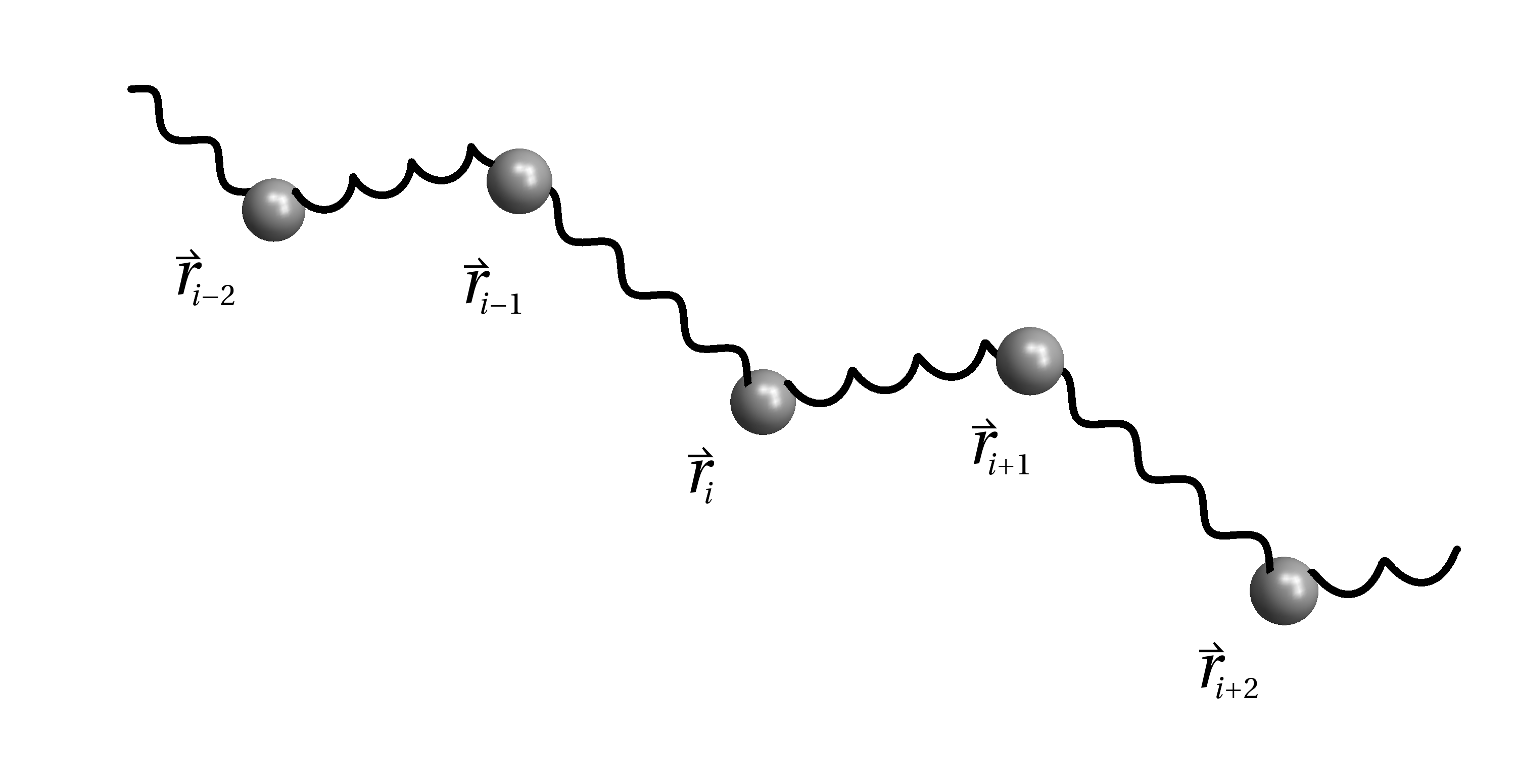}
  \caption{Artistic illustration of a bead-spring model of a
    polymer. In addition to the harmonic confinement between the
    beads, an angular confinement is also implemented.}
  \label{fig:polymer}
\end{figure}

In this section, we investigate the persistence probability of an
interacting chain of $N$ Brownian particles, in a two-dimensional
planar solvent subjected to the stationary flow of
\fref{eq:velprf}. In addition to the harmonic confinement
$U(r)=\frac{1}{2} \kappa (r - R_0)^2$ of a particle to its nearest
neighbors, we impose a bending potential
\begin{equation}
  \label{eq:angular_cnf}
  U^{\;\textrm{bend}}(\theta)=\sum_{j=2}^{N-1}\frac{\kappa_\theta}{2}
  (1+\vec{\hat{u}}_j \vec{\hat{u}}_{j-1})^2,
\end{equation}
where $\vec{\hat{u}}_i = \vec r_{i+1} -\vec r_i/|\vec r_{i+1} -\vec
r_i|$. This discrete model of a chain of Brownian particle is often
used in simulations to model semi-flexible polymer chains
\cite{Pham2008,Manghi2006}. In the absence of the bending potential,
the model corresponds to that of a Rouse chain and the persistence
probability of such a chain in a simple shear flow has been studied by
Bhattacharya \etal \cite{Bhattacharya2007}.  The resulting expression
for the bending force takes the form,
\begin{equation}
  \label{eq:bending_force}
  \vec f^{\;\textrm{bend}}_i = -\kappa_{\theta}\sum_{j=2}^{N-1}
  (1+\vec{\hat{u}}_j 
  \vec{\hat{u}}_{j-1}) \dfrac{\partial
  }{\partial \vec r_i}\vec{\hat{u}}_j\vec{\hat{u}}_{j-1}
\end{equation}
Using the explicit form of $\vec{\hat{u}}_i$, and assuming that the
confinement strength of the harmonic force and the bending
rigidity is sufficiently large so that we are in weakly bending limit,
such that $|\vec{\hat{u}}(i+1)-\vec{\hat{u}}(i)| \ll 1 $ \cite{Morse1998},
the force on particle $i$ becomes,
\begin{equation}
  \label{eq:bending_force_explicit}
  \vec f^{\;\textrm{bend}}_i=-2 \frac{\kappa_{\theta}}{R_0^4} 
  (-\vec r_{i-2}+4 \vec r_{i-1}-6 \vec r_i
  +4 \vec r_{i+1}-\vec r_{i+2} ),
\end{equation}
with $R_0$ as the inter-particle separation. Note that in this
approximation, the force due to the harmonic confinement can be
neglected.  The equation of a Brownian particle then becomes,
\begin{align}
  \label{eq:langevin_chain}
\begin{split}
  \frac{\upd x_i}{\upd t}&=-2 \frac{\kappa_{\theta}}{R_0^4}
  (-x_{i-2} + 4x_{i-1}-6x_i +4x_{i+1}-x_{i+2} )+\eta_x(i,t)\\
  \frac{\upd y_i}{\upd t}&=-2 \frac{\kappa_{\theta}}{R_0^4}
  (-y_{i-2} + 4y_{i-1}-6y_i +4y_{i+1}-y_{i+2} )+ a x_i(t)+\eta_y(i,t)
\end{split}
\end{align}
The above equations are only valid for $N-4$
particles, and the $4$ particles at the boundary will have different
equations of motion. However, for an infinitely long chain, the
monomer dynamics at late times is not sensitive to the boundary
conditions and accordingly a continuum version of the equations can be
formulated by replacing $i$ with a continuous variable $s$,
\begin{align}
  \label{eq:langevin_chain_continuous}
\begin{split}
  \frac{\partial }{\partial t}\!\! \!\bfrac{x(s,t)}{}
  &=-\bar{\kappa}\;\frac{\partial^4}{\partial s^4}\!\!\!\bfrac{
    x(s,t)}{}+\eta_x(s,t)\\
  \\
  \frac{\partial}{\partial t}\!\!\! \bfrac{y(s,t)}{}
  &=-\bar{\kappa}\;\frac{\partial^4}{\partial s^4}\!\!\!\bfrac{
    y(s,t)}{}+ax(t)+\eta_y(s,t)
\end{split}
\end{align}
Fourier transforming the above equations leads to,
\begin{equation}
  \label{eq:langevin_ft_x}
  \frac{\partial\tilde{x}}{\partial t}=-\bar{\kappa}\; k^4\;
  \tilde{x}(k,t)+\tilde{\eta}_x(k,t)\\
\end{equation}
\begin{equation}
  \label{eq:langevin_ft_y}
  \frac{\partial\tilde{y}}{\partial t}=-\bar{\kappa}\; k^4\;
  \tilde{y}(k,t)+a \tilde{x}(k,t)
\end{equation}
We have dropped the stochastic noise term in \fref{eq:langevin_ft_y},
since at late-times the second term $a x(t)$ becomes the dominant term
compared to the stochastic force. The formal solutions of
\fref{eq:langevin_ft_x} and \fref{eq:langevin_ft_y} are given by,
\begin{align}
  \label{eq:solution_x_and_y}
 \tilde{x}(k,t)=\int_0^t\; \upd t'\; e^{-\bar{\kappa}\;k^4\;(t-t')}
 \;\tilde{\eta}_x(k,t') \\
 \tilde{y}(k,t)=\int_0^t\; \upd t'\; e^{-\bar{\kappa}\;k^4\;(t-t')}\; 
 \tilde{x}(k,t'). 
\end{align}
Using the noise correlation for $\la
\tilde{\eta}_x(k,t)\tilde{\eta}_x(k',t) \ra=
2D\delta(k+k')\delta(t-t')$, the two-time correlation function for
$\tilde{x}$ reads,
\begin{align}
  \label{eq:two_time_x}
  \nonumber
  \la \tilde{x}(k,t_1)\tilde{x}(k',t_2) \ra&=
  \frac{D}{\bar{D}(k)} \delta(k+k')
  e^{-\bar{D}(k)(t_1+t_2)} \left[e^{2 \bar{D}(k) \min (t_1,t_2)}-1 \right]\\
  &=\frac{D}{\bar{D}(k)} \delta(k+k') \left[e^{- \bar{D}(k) (t_1-t_2)}
   -e^{-\bar{D}(k)(t_1+t_2)}
  \right],
\end{align}
where $\bar{D}(k)=\bar{\kappa}k^4$ is the diffusion coefficient of a
mode $k$. The second line of \fref{eq:two_time_x} assumes $t_1>t_2$.
Similarly, the two-time correlation function for the $y(s,t)$ can be
constructed from \fref{eq:two_time_x},
\begin{multline}
  \label{eq:two_time_y}
  \la y(s,t_1) y(s,t_2) \ra= a^2 \int \frac{\upd k_1}{2 \pi} \int
  \frac{\upd k_2}{2 \pi} e^{-(k_1+k_2)s} \int_0^{t_1} \upd t'_1
  \int_0^{t_2} \upd t'_2 \\
  e^{-\bar{D}(k_1) (t_1-t'_1)}e^{-\bar{D}(k_2)(t_2-t'_2)}
 \la \tilde{y}(k_1,t'_1) \tilde{y}(k_2,t'_2) \ra
\end{multline}
Substituting \fref{eq:two_time_x} in the equation above and
subsequently integrating over the delta function and renaming the
dummy variable $k_1$ we arrive at,
\begin{multline}
  \label{eq:two_time_y1}
  \la y(s,t_1) y(s,t_2) \ra= a^2 \int \frac{\upd k}{2 \pi}
  \int_0^{t_1} \upd t'_1 \int_0^{t_2} \upd t'_2 \;e^{-\bar{D}(k)
    (t_1+t_2)}\; \\
   \frac{D}{\bar{D}(k)}\; \left[e^{2
      \bar{D}(k) \min(t'_1,t'_2)}-1\right]\\
  = \frac{a^2D}{\bar{\kappa}}\int_0^{t_1} \upd t'_1 \int_0^{t_2} \upd t'_2
  \left[\int \frac{\upd k}{2 \pi}\frac{1-e^{-\bar{\kappa}
        k^4(t_1+t_2)}}{k^4} \right.\\
      \left. -\int \frac{\upd k}{2
      \pi}\frac{1-e^{-\bar{\kappa} k^4(t_1+t_2-2\min(t'_1,t'_2))}}{k^4}\right].
\end{multline}
The integral over $k$ yields,
\begin{multline}
  \label{eq:two_time_y2}
  \la y(s,t_1) y(s,t_2) \ra= \frac{a^2 D}{\bar{\kappa}}  \int_0^{t_1} \upd t'_1
  \int_0^{t_2} \upd t'_2 \; \left[(t_1+t_2)^{\frac{3}{4}} \right.\\
  \left. - (t_1+t_2-2 \min(t'_1,t'_2))^{\frac{3}{4}} \right]
\end{multline}
The integral over the $t'_1$ and $t'_2$ finally yields,
\begin{multline}
  \label{eq:two_time_y3}
  \la y(s,t_1) y(s,t_2) \ra= \frac{2a^2 D
    \Gamma(\tfrac{1}{4})}{3\bar{\kappa}} 
  \left[t_1 t_2 (t_1+t_2)^{\frac{3}{4}}-\frac{2}{11}
    (t_1+t_2)^{\frac{11}{4}} \right.\\
\left. + \frac{2}{11}(t_1-t_2)^{\frac{11}{4}}\right]
\end{multline}
Note that the correlation function is independent of $s$, and this
translational invariance is a consequence of taking the limit of an
infinitely long chain. We shall, henceforth, drop the continuous
variable from the notations. At this point, we justify dropping the
stochastic noise term in \fref{eq:langevin_ft_y}. The inclusion of
this term leads to an additional term $\mathcal{O}(t^{3/4})$, which
can be neglected compared to the terms in correlation functions in
\fref{eq:two_time_y3}, which are of $\mathcal{O}(t^{11/4})$.

\begin{figure}
  \includegraphics[width=\linewidth]{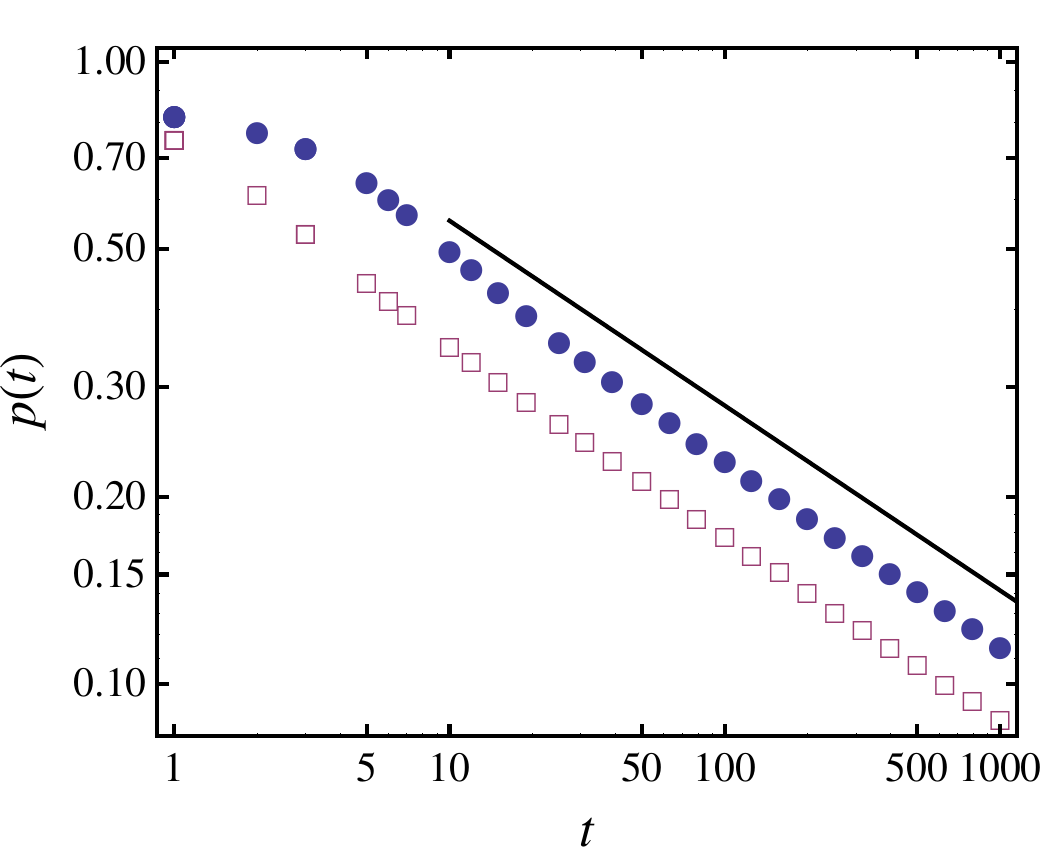}
  \caption{Persistence probability of a semi-flexible polymer in a
    shear flow for values of $a=0.1$ ({\color{myblue}
      {\large $\bullet$}}) and $0.50$ ({\color{mypurple}
      $\square$}). The solid black line is a plot of $t^{-\theta}$,
    with $\theta=0.29695$, the prediction of IIA. A power law fit to
    the asymptotic of the data yields $\theta=0.306236$ for $a=0.1$
    and $\theta=0.297205$ for $a=0.5$. }
  \label{fig:perprob_semi}
\end{figure}

The transformation to a stationary process is done using the identical
procedure as outlined in the previous sections. Defining the
normalized variable $\overline{Y}=y(t)/\sqrt{\la y^2(t) \ra}$ and the
transformation $e^T=t$ yields a stationary correlation function,
\begin{multline}
  \label{eq:two_time_y_st}
  C(T)\equiv \la \overline{Y}(0)\overline{Y}(T)\ra=\frac{11}{3}
   \left[\cosh\left(\frac{T}{2}\right)\right]^{\frac{3}{4}}-\frac{8}{3}
   \left[\cosh\left(\frac{T}{2}\right)\right]^{\frac{11}{4}}\\
   +\frac{8}{3}
   \left[\sinh\left(\frac{T}{2}\right)\right]^{\frac{11}{4}}.
\end{multline}
Since the correlation function is not exponentially decaying, the
simple route to determine the persistence probability can not applied
here. A Taylor expansion of $C(T)$ in the neighborhood of zero gives,
\begin{equation}
  \label{eq:taylor_expansion}
  C(T)=1-\frac{55}{96}T^2 +\mathcal{O}(T^{11/4}),
\end{equation}
demonstrating that $\overline{Y}$ is a smooth process with a finite
number of zero crossings. The density of zero crossings is given
by \cite{Slepian1962,Rice1945},
$\rho\equiv \sqrt{-C''(0)}/\pi=\frac{1}{\pi}\sqrt{\frac{55}{48 }}$ and the
mean interval size $\la T \ra$ is given by $\la T \ra=1/\rho$. To
determine the exponent $\theta$ we use the Independent Interval
approximation \cite{Majumdar1996}, where the intervals between
successive zero crossings are assumed to be independent. The
distribution $P(T)$ of the intervals between the successive zeros can
be related to the stationary correlator $C(T)$ in the Laplace domain
as \cite{Bhattacharya2007},
\begin{equation}
  \label{eq:distribution_of_zeros}
  \tilde{P}(s)=\frac{1-\tfrac{\la T \ra s}{2} (1-s
    \tilde{C}(s))}{1+\tfrac{\la T \ra s}{2} (1-s \tilde{C}(s))},
\end{equation}
where the tildes refer to the Laplace transforms of the corresponding
quantities. Finally, the estimation of the exponent $\theta$ from the
interval size distribution translates to the determination of the
simple pole of $\tilde{P}(s)$, that is, the root of the denominator in
\fref{eq:distribution_of_zeros} \cite{Bhattacharya2007,Majumdar1996},
\begin{multline}
  \label{eq:denominator}
  F(s)=1+\frac{\la T \ra s}{2}\left[1-s \int_0^\infty \upd T
    e^{-sT}\left(\frac{11}{3}
   \left[\cosh\left(\frac{T}{2}\right)\right]^{\frac{3}{4}} \right.\right.\\
   \left.\left. -\frac{8}{3}
   \left[\cosh\left(\frac{T}{2}\right)\right]^{\frac{11}{4}}
   +\frac{8}{3}\left[\sinh\left(\frac{T}{2}\right)\right]^{\frac{11}{4}}
   \right) \right]
\end{multline}
The numerical estimation of the root of $F(s)$ yields a value of
$\theta=0.29695$. To verify this result, we performed numerical
integration of \fref{eq:langevin_chain_continuous} by discretizing the
equation on a lattice with $1024$ lattice points,
\begin{multline}
  \label{eq:discrete_langevin}
  x_i(t_{m+1})=x_i(t_m)-\Delta t \bar{\kappa} \left[-x_{i-2}(t_m)+4
    x_{i-1}(t_m) -6x_i(t_m) \right. \\
  \left. +4 x_{i+1}(t_m)-x_{i+2}(t_m)\right]+\sqrt{2 D \Delta t}
  \mathcal{N}_i(t_m) \\
  y_i(t_{m+1})=y_i(t_m)-\Delta t \bar{\kappa} \left[-y_{i-2}(t_m)+4
    y_{i-1}(t_m) -6y_i(t_m) \right. \\
  \left. +4 y_{i+1}(t_m)-y_{i+2}(t_m)\right]+a x_i(t_m),
\end{multline}
where $\mathcal{N}_i(t_m)$ is a Gaussian random number with zero mean
and unit variance and $\Delta t=0.01$ is the integration time
step. The free boundary conditions were implemented by choosing
$\vec{r}(0)=\vec{r}(1)=\vec{r}(2)$ and
$\vec{r}(N)=\vec{r}(N-1)=\vec{r}(N-2)$. In the course of the
simulation, the sign of the variable $y_i(t)$ were monitored and the
persistence probability was defined as the fraction of the lattice
points for which the sign of $y_i(t_m)$ was same as that of $y_i(0)$.
Further, the measured persistence probability was averaged over $500$
independent configurations. The asymptotic of the probability decays
as a power law with an exponent close to the prediction of IIA (see
\fref{fig:perprob_semi}).

\section{Conclusion}
In conclusion, we have presented results for the two-time correlation
functions of a single free and confined Brownian particle in a simple
shear flow. The confinement of the Brownian particle is modeled as a
harmonic trap, often encountered in trapping and tracking
experiments. The persistence probability, defined as the probability
that the sign of the stochastic observable has not changed sign up to
time $t$ is constructed from the correlation function. The probability
shows two distinct algebraic decays. For short times $t<<a^{-1}$, when
the particle does not feel the effect of the shear, the motion is
found to be purely diffusive while at late times the motion is super
ballistic with the mean-square-displacement along the direction of
shear scaling as $t^3$. We have also extended the analysis to a chain
of Brownian particles interacting via a harmonic potential and a
bending potential. The asymptotic of the persistence probability is
found to decay as a power law with as exponent in close agreement with
that of the predictions from Independent Interval Approximations.

\bibliography{library}
\bibliographystyle{epj}
\end{document}